\documentclass[a4paper,11pt]{article}

\usepackage{graphicx}
\usepackage{verbatim}
\usepackage{amsmath,amsthm,amssymb,amsfonts,color}
\usepackage{float}
\usepackage{enumitem}

\DeclareGraphicsExtensions{.eps,.ps}   
\usepackage{pos}

\title{Toward a novel determination of the strong QCD coupling at the Z-pole}

\author*[a]{Chik Him Wong}
\author[a]{Szabolcs Borsanyi}
\author[a,d,e,f]{Zoltan Fodor}
\author[b]{Kieran Holland}
\author[c]{Julius Kuti}

\affiliation[a]{University of Wuppertal,\\
Gaussstrasse 20, 42119 Wuppertal,Germany}

\affiliation[b]{University of the Pacific,\\
3601 Pacific Ave., Stockton, CA 95211, USA}

\affiliation[c]{UC, San Diego,\\
9500 Gilman Drive, La Jolla, CA 92093-0021, USA}

\affiliation[d]{ELTE E\"{o}tv\"{o}s Lor\'{a}nd University, Institute for Theoretical Physics,\\
P\'{a}zm\'{a}ny P\'{e}ter s\'{e}t\'{a}ny 1/A,H-1117, Budapest, Hungary}

\affiliation[e]{Pennsylvania State University,\\
State College, Pennsylvania 16801,USA}

\affiliation[f]{J\"{u}lich Supercomputing Centre, Forschungszentrum J\"{u}lich,\\
D-52425 J\"{u}lich, Germany}

\emailAdd{cwong@uni-wuppertal.de}
\emailAdd{borsanyi@uni-wuppertal.de}
\emailAdd{fodor@bodri.elte.hu}
\emailAdd{kholland@PACIFIC.EDU}
\emailAdd{jkuti@ucsd.edu}

\abstract{
We test here our recently introduced new lattice method  for the $\beta$-function defined  over  infinite Euclidean space-time in the continuum from scale changes generated by infinitesimal or finite steps of the renormalized gauge coupling on the gradient flow. Harlander and Neumann calculated in this scheme the three-loop approximation to the continuum $\beta$-function. Our goal is the nonperturbative lattice implementation of the scheme which we tested originally in the chiral limit of the sextet model and in multi-flavor QCD with ten and twelve flavors of massless fermions.  Results are reported here in the SU(3) Yang-Mills gauge sector without dynamical fermions and 
in ten-flavor QCD with massless femions. The three-loop gradient flow based $\beta$-function of Harlander and Neumann is used  to connect the  $\Lambda_{\overline{\rm MS}}$ scale of the SU(3) Yang-Mills gauge theory with the  nonperturbative flow time scale $t_0$, or the equivalent Sommer scale $r_0$.  Similarly, the $\Lambda_{\overline{\rm MS}}$ scale is connected with a selected nonperturbative scale  in the ten-flavor theory, a pilot study of our new lattice based nonperturbative $\beta$-function for high precision determination of the strong coupling $\alpha_s$ at the Z-boson pole in QCD with three massless fermion flavors. This goal  is an important alternative to results from the finite volume based step $\beta$-function of the Alpha collaboration. Work is ongoing on direct application of the method to  QCD with three massless fermion flavors.
}

\FullConference{%
  The 39th International Symposium on Lattice Field Theory (Lattice2022),\\
  8-13 August, 2022 \\
  Bonn, Germany 
}

\begin{document}
\maketitle
\section{ Introduction }
	
Earlier, we have developed high precision methods to determine step $\beta$-functions of important gauge theories defined  over  finite physical volumes in the continuum from stepped scale changes of the renormalized gauge coupling on the gradient flow. The scale changes were controlled by the linear size of the physical volume. Conformal and near-conformal gauge field theories close to the lower edge of the conformal window were the primary targets of the earlier studies which remain important directions in lattice BSM investigations. We use periodic gauge field boundary conditions in the finite volume where the zero momentum mode of the gauge field is not removed from the renormalization procedure.  The presence of the zero mode in finite volume leads to known complications of the perturbative expansion of the $\beta$-function which is only straightfoward to one-loop order in this case. Consequently, the connection between the  $\Lambda_{\overline{\rm MS}}$ scale and the target nonperturbative scale is less controlled and more prone to systematic errors in any application. In contrast, the  infinite volume based $\beta$-function is known to three-loop order making this connection much better controlled, leading us to the change in strategy to {\em infinite} volume, as reported here. Although not discussed here, dealing with the zero mode complications as the alternate possibility of determining the 3-loop step $\beta$-function in the continuum over {\em finite} physical volume did not escape our attention.

We report here some new results from the infinite volume based $\beta$-function  strategy, testing  our recently introduced  lattice algorithms  in the continuum limit from scale changes of infinitesimal or finite steps in the renormalized gauge coupling on the gradient flow. Harlander and Neumann calculated in this scheme the three-loop approximation to the continuum $\beta$-function~\cite{Harlander:2016vzb} with extended follow-up in~\cite{Artz:2019bpr}. Our goal is the new nonperturbative lattice implementation of the scheme. 

The infinite-volume $\beta$-function is based directly on the gradient flow coupling $g^2(t)$ defined in infinite volume as a function of  continuous flow time $t$, probing the theory at the energy scale $\mu$ given by $\mu = 1/\sqrt{8 t}$  af flow time $t$. This allows the direct definition  $\beta(g^2(t))=t \cdot dg^2/dt$ from infinitesimal RG scale change using the opposite of the standard sign convention, inherited from common practice in lattice BSM studies. Flow time $t$ at fixed lattice spacing $a$  is always expressed dimensionless  as $t/a^2$. In the massless fermion limit we have to take limits from finite lattice volumes to the infinite lattice volume at fixed bare gauge coupling and we also have to take the $a\rightarrow 0$ limit of  $\beta(g^2(t))=t \cdot dg^2/dt$ in infinite lattice volume at fixed renormalized gauge coupling. In ordering the limits, there are several lattice algorithms to implement the target goals of the $a\rightarrow 0$ continuum  limit from massless fermions in finite lattice volumes while reaching the infinite physical volume. 
There are also algorithms which can reach the massless fermion limit over infinite volumes from small fermion mass deformations in the $\epsilon$-regime, or in the $p$-regime of chiral perturbation theory if the theory has a phase with chiral symmetry breaking. In fact, even mass-dependent $\beta$-functions could be probed over infinite continuum volumes in future applications.

We tested earlier two paths (algorithms) for the new $\beta$-function defined in  infinite continuum volume with massless fermions~\cite{Fodor:2017die,Fodor:2019ypi}. Originally, we introduced the new algorithm to match scales set by finite-volume step $\beta$-functions in massless near-conformal gauge theories to scales set by the infinite-volume $\beta$-function in the chiral limit of fermion mass deformations reached from the phase with spontaneous chiral symmetry breaking. The pilot study of this algorithm was the near-conformal two-flavor sextet model reaching the massles chiral limit from small fermion mass deformations $m$ of the p-regime in the chiral symmetry breaking phase~\cite{Fodor:2017die}. An alternative implementation of the infinite-volume $\beta$-function through $t \cdot dg^2/dt$ is applied here using simulations with massless fermions set  in finite lattice volumes and extrapolated to the infinite lattice volume limit at fixed lattice spacing. Tests are reported here in SU(3) Yang-Mills gauge theory and the ten-flavor QCD model based on simulations directly at $m = 0$.  The infinite-volume limit is taken at fixed reference values of the gradient flow time $t/a^2$  before the cutoff $a$ is eliminated in  the $a^2/t \rightarrow 0$  continuum limit. This lattice algorithm was tested earlier in the ten-flavor and twelve-flavor BSM theories~\cite{Fodor:2019ypi} and  in the two-flavor QCD model~\cite{Hasenfratz:2019hpg} . 

\section{Outline}
\vskip -0.1in

Test results are reported in Section 3  for the SU(3) Yang-Mills gauge sector of QCD without dynamical fermions. 
 The three-loop $\beta$-function of Harlander and Neumann is used in the gradient flow based renormalization scheme to connect the  $\Lambda_{\overline{\rm MS}}$ scale of the SU(3) Yang-Mills gauge sector with the  nonperturbative flow time scale $t_0$, or the equivalent Sommer scale $r_0$. Similarly, in Section 4, the $\Lambda_{\overline{\rm MS}}$ scale is connected with a selected nonperturbative scale  in the ten-flavor theory. The two tests  are  pilot studies in  applying the new lattice based nonperturbative $\beta$-function to high precision determination of the strong coupling $\alpha_s(m_Z)$ at the Z-boson pole in QCD with three massless fermion flavors. This goal is an important alternative to results from the finite volume based step $\beta$-function of the Alpha collaboration. Work, not reported here, is ongoing on application of the method directly to  QCD with three massless fermion flavors.

\section{Precision tests of the Yang-Mills sector of QCD without dynamical fermions}
\vskip -0.1in

\begin{figure}
	\begin{center}
        \scalebox{0.3}{\includegraphics{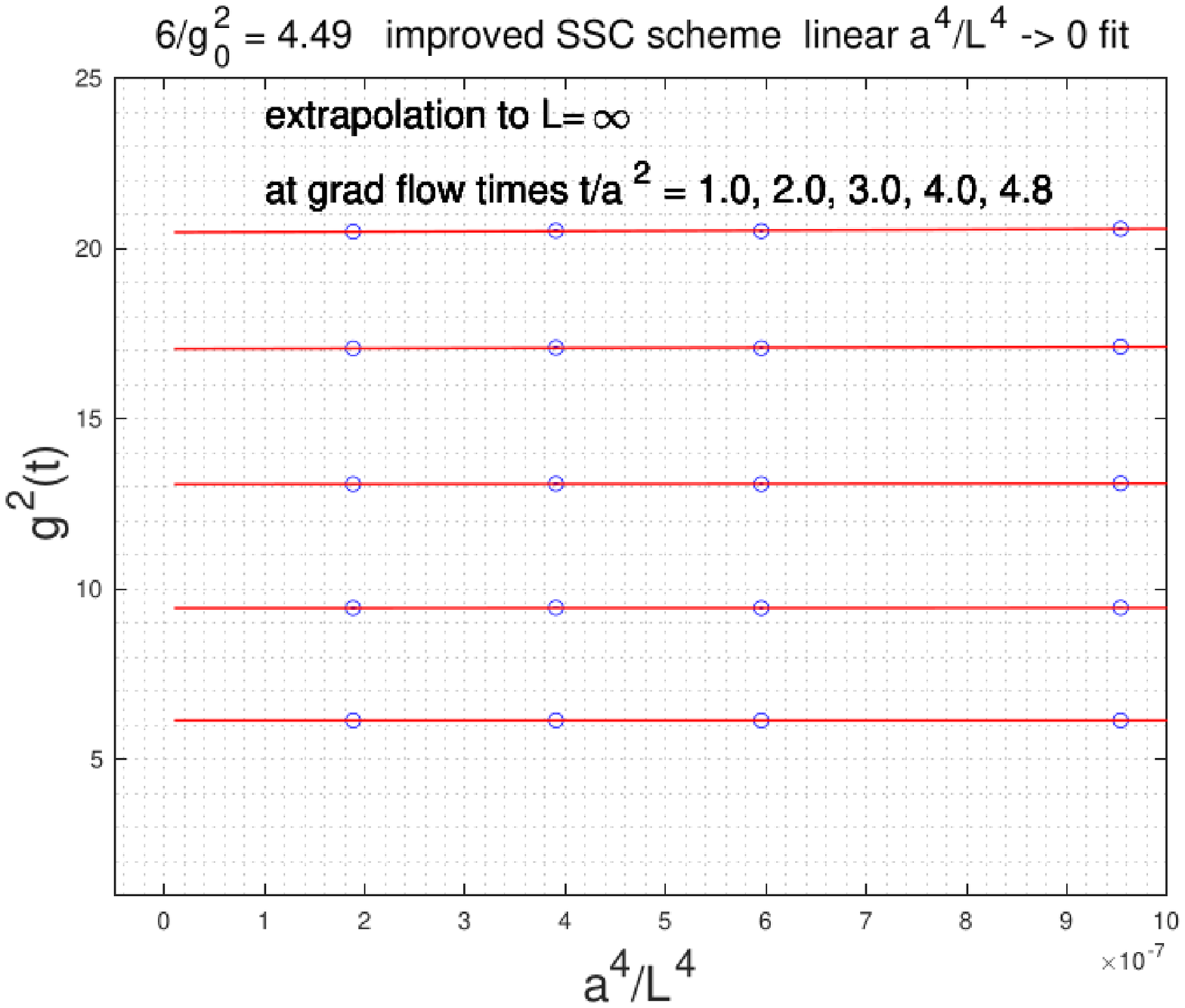}}
        \scalebox{0.3}{\includegraphics{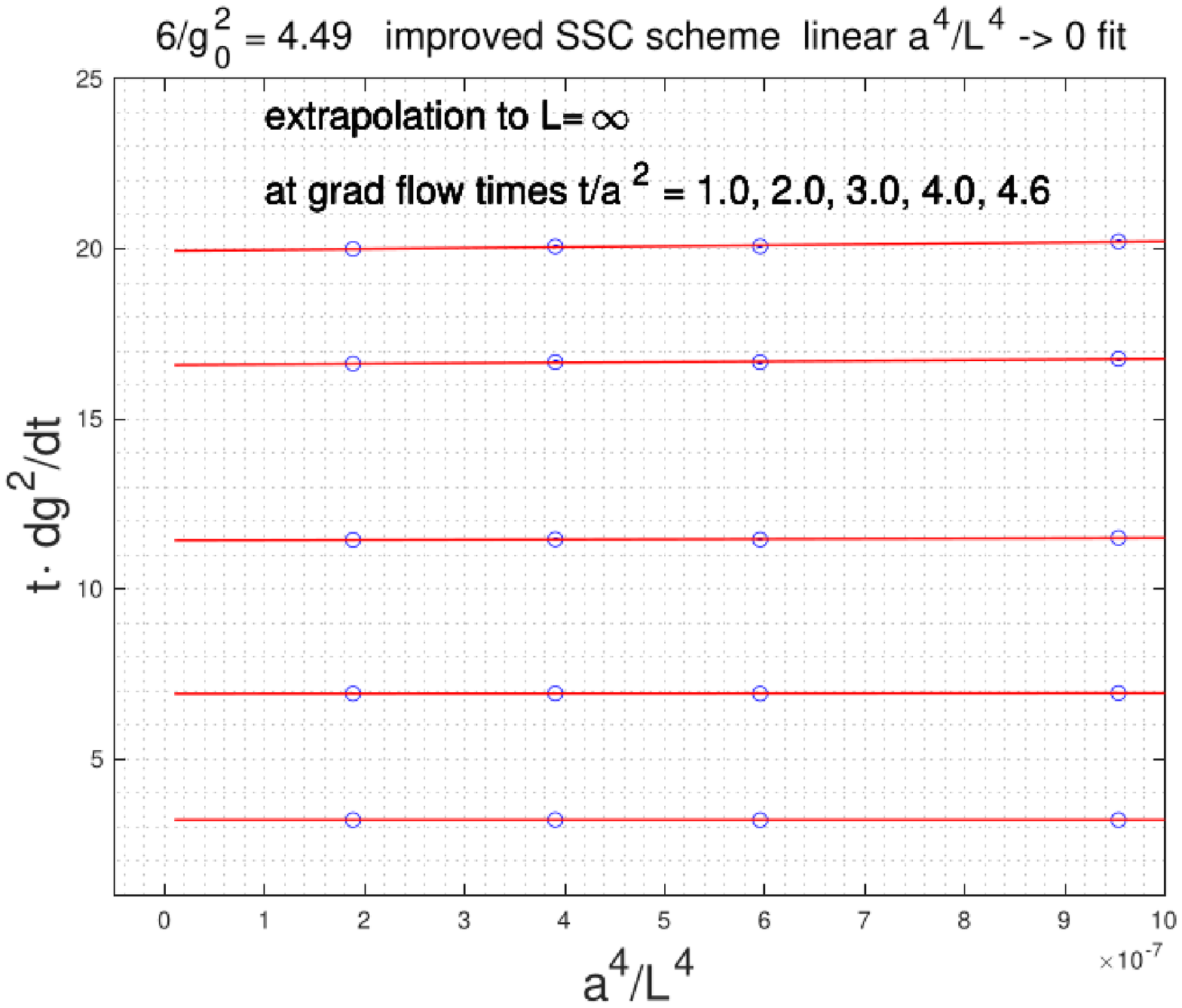}}
	\end{center}		
	\vskip -20pt
	\caption{\label{YM1}  {\small Step 1 of the SU(3) Yang-Mills analysis is shown at some selected values of the flow time $t/a^2$. }}
\end{figure}

\begin{figure}	
	\begin{center}
        \scalebox{0.3}{\includegraphics{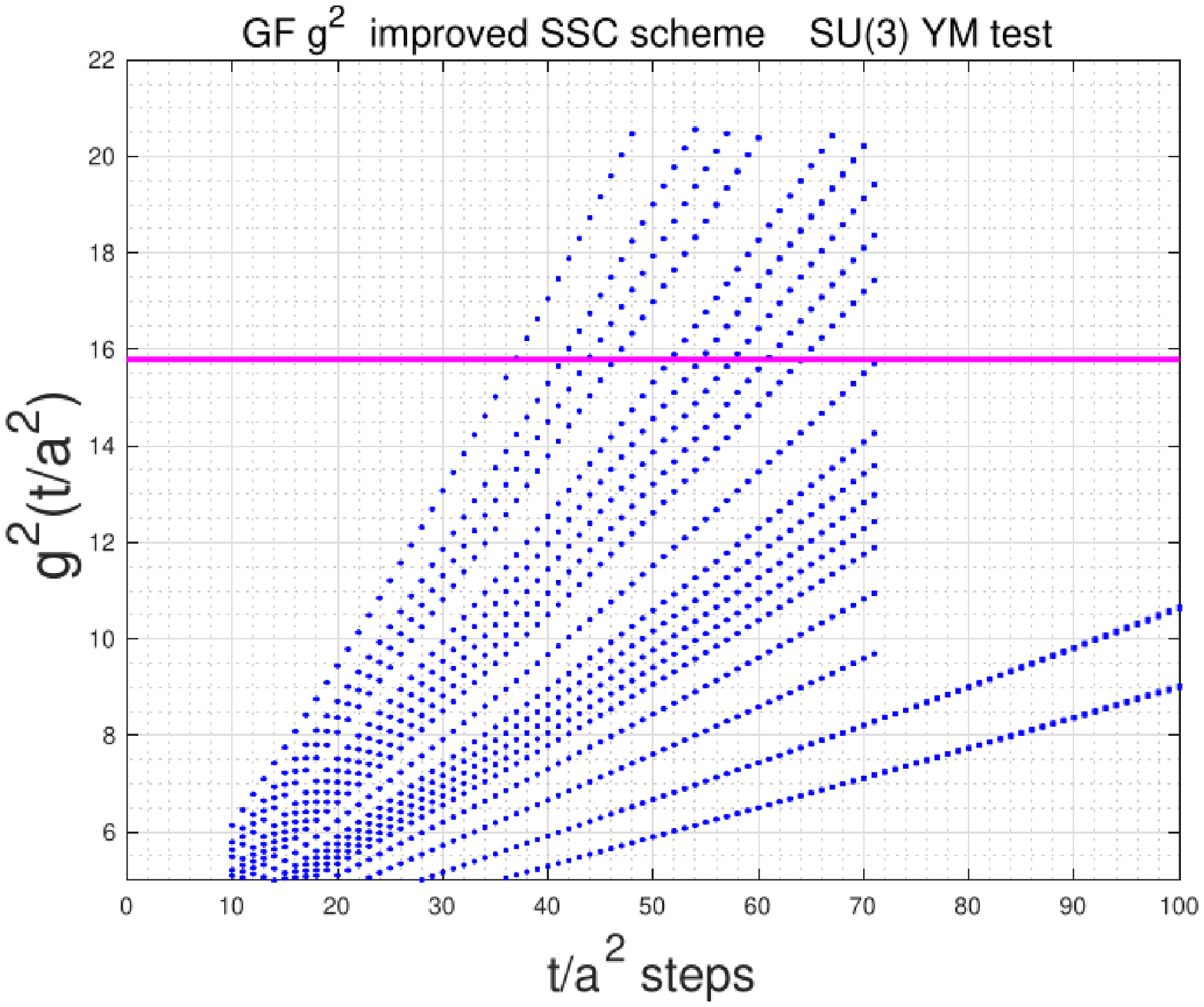}}
        \scalebox{0.3}{\includegraphics{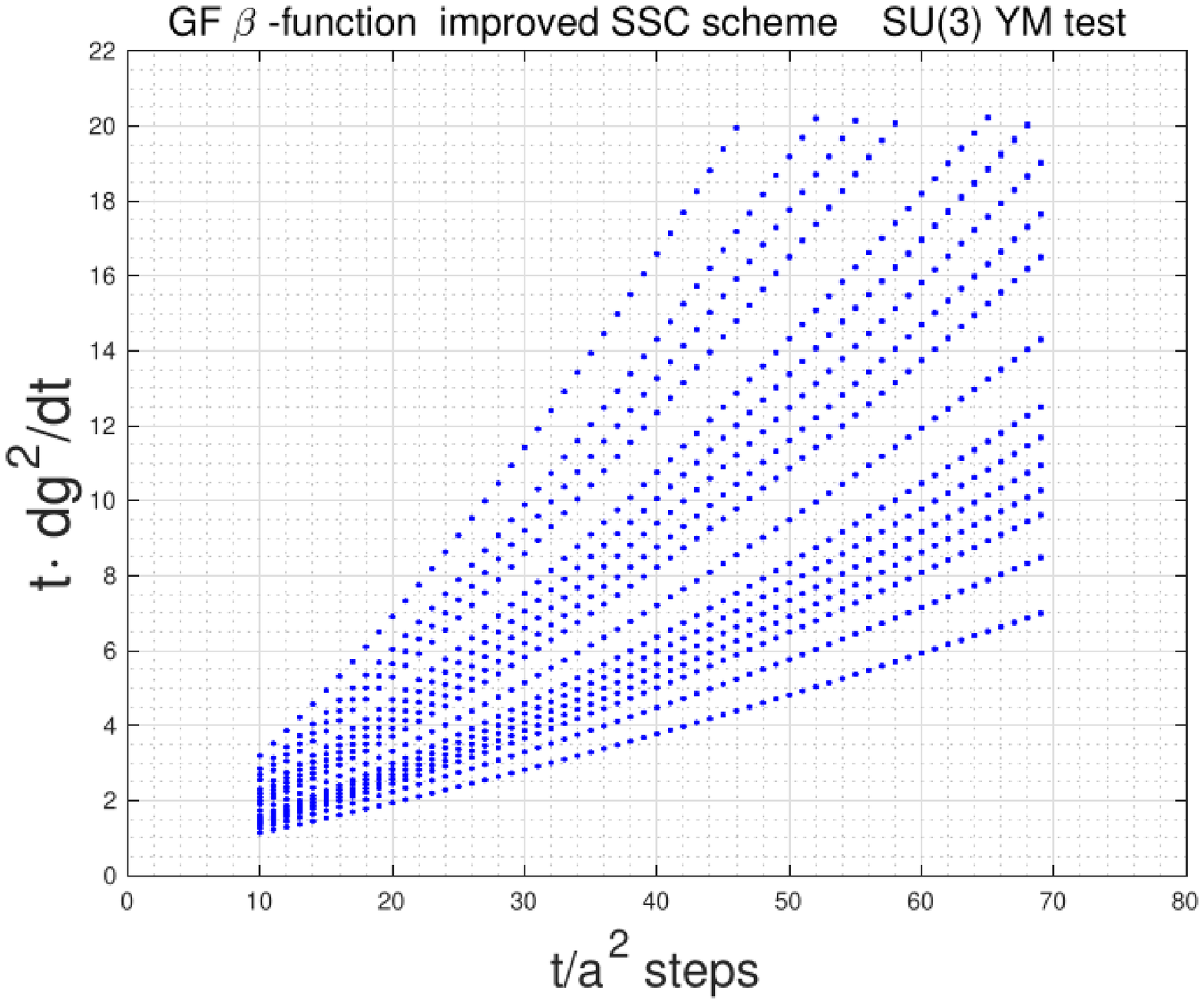}}
	\end{center}		
	\vskip -20pt
	\caption{\label{YM2}  {\small  The right panel shows the extrapolated infinite-volume  $\beta=t dg^2/dt$ while the renormalized coupling $g^2$ is held fixed on the left panel at some target value.  The steps are counted in units of the step $\epsilon$.}}
\end{figure}

\begin{figure}
	\begin{center}
        \scalebox{0.3}{\includegraphics{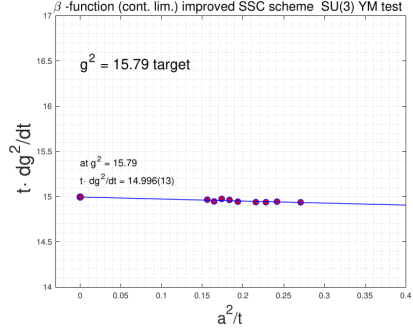}}
	\end{center}		
\vskip -0.2in
	\caption{\label{YM3}  {\small Step 3 of the SU(3) Yang-Mills analysis with  $t\cdot dg^2/dt$ extrapolated in $a^2/t$ 
			to the continuum limit at $g^2=15.79$ with per mille accuracy.}}
\end{figure}

Algorithmic implementation of the infinite-volume $\beta$-function through $t \cdot dg^2/dt$ is applied in this section to new tests in the SU(3) Yang-Mills gauge sector of QCD without dynamical fermions. 
The infinite-volume limit is taken at fixed reference values of the gradient flow time $t/a^2$ in lattice units $a$ before the cutoff $a$ is eliminated in  the $a^2/t \rightarrow 0$  limit. 
To demonstrate achievable high precision, we present an efficient and accurate  implementation of the algorithm where 
the derivative $dg^2/dt$  with respect to  the flow time variable $t/a^2$ is approximated numerically by  five-point discretization in the small flow time step $\epsilon$,
\vskip -0.1in
\begin{equation}\label{derivative}
[-g^2(t + 2\epsilon) + 8g^2(t + \epsilon) - 8g^2(t - \epsilon) + g^2(t - 2 \epsilon)]/(12 \epsilon) = dg^2/dt + {\mathcal O}(\epsilon^4)~.
\end{equation}
\noindent{The} discrete flow time step $\epsilon$ in Eq.~(\ref{derivative}) is used in an adaptive  integration scheme of the gradient flow equations on the lattice with the required accuracy level. For cross-checks, we also use in the lattice analysis spline based determination of the $\beta$-function $t \cdot dg^2/dt$  and a robust interpolation method of derivatives from a scheme introduced by  Akima~\cite{Akima:1970}.
\vskip 0.05 in
\noindent{The algorithm of the lattice analysis has three steps:}
\vskip 0.05in
\noindent{\bf Step 1:}  
The first step  is applied to a large set of lattice ensembles in the broad range of 39  bare gauge couplings $6/g_0^2$ at linear volume sizes $L=32,36,40,48,64$  with $L=80$ and $L=96$ added for cross checks at a few select couplings $6/g_0^2 = 4.81, 6.36, 7.8$. At each $L$ and at each of the 39  bare gauge couplings $6/g_0^2$ we measure the renormalized coupling $g^2(t)$ and the $\beta$-function defined as $t\cdot dg^2/dt$ over discretised small step increments $\epsilon$ of the  gradient flow time $t/a^2$, as given in Eq.~(1). At any given value of $6/g_0^2$ for any given finite flow time step we extrapolate the volume set $L=32,36,40,48,64$  to the $L\rightarrow \infty$ limit.This now defines $g^2(t)$ and $t\cdot dg^2/dt$ in the infinite lattice volume limit at each flow time step $\epsilon$  and at each lattice spacing set by  $6/g_0^2$. Step 1  is illustrated in Fig.~\ref{YM1} with improved SSS, SSC, WSC, and WSS  schemes defined in~\cite{Fodor:2022qfe}. 
\vskip 0.05in
\noindent{\bf Step 2:} We select targeted continuum $g^2(t)$ values, like $g^2(t) = 15.79$, or any other  selected value, implicitly defining the continuum flow time scale in physical units although not expressed yet in terms of the scale $\Lambda_{\overline{\rm MS}}$. We know for example that $g^2(t) = 15.79$ would define implicitly the $t_0$ continuum scale~\cite{Luscher:2010iy} to be expressed in  $\Lambda_{\overline{\rm MS}}$ units as the set goal. Since we know the flow time values $t/a^2$ in cutoff units for each step $\epsilon$, for each $g^2$ selection we can read off  the values of $t\cdot dg^2/dt$ at each $t/a^2$ by minimal interpolation within the size of a flow time step $\epsilon$. The small error of the interpolation  is included in the analysis. Step 2 is illustrated in Fig.~\ref{YM2}. 
\vskip 0.05in
\noindent{\bf Step 3:}  We extrapolate the values of $t\cdot dg^2/dt$ from matching flow time  values $t/a^2$  to the continuum limit $a^2/t \rightarrow 0$ at  the chosen target $g^2(t)$ of the continuum theory. Leading order linear fit in  $a^2/t$ is shown in Fig.~\ref{YM3} for the selection of $g^2(t) = 15.79$ .

\begin{figure}	
	\begin{center}
        \scalebox{0.25}{\includegraphics{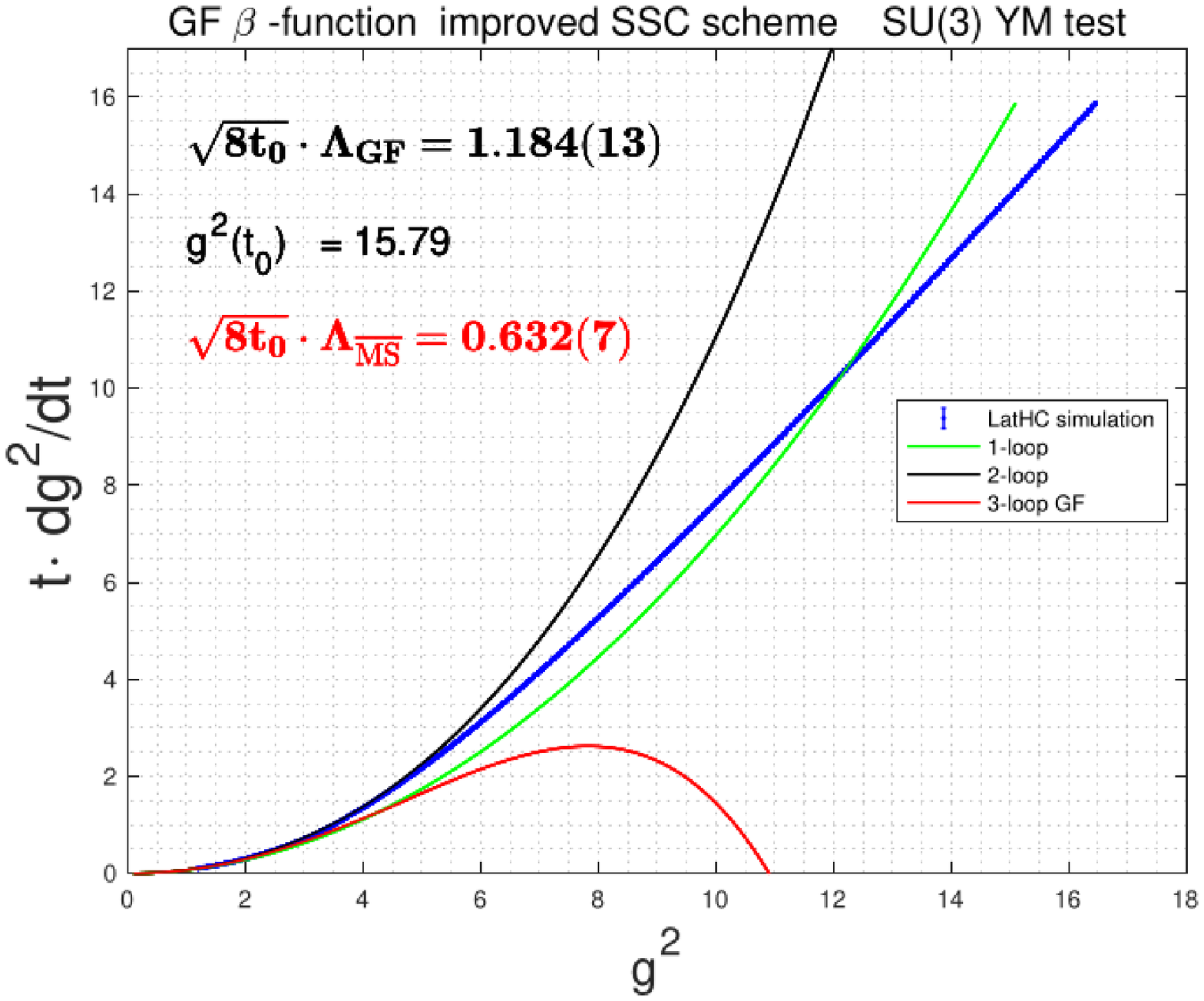}}
        \scalebox{0.25}{\includegraphics{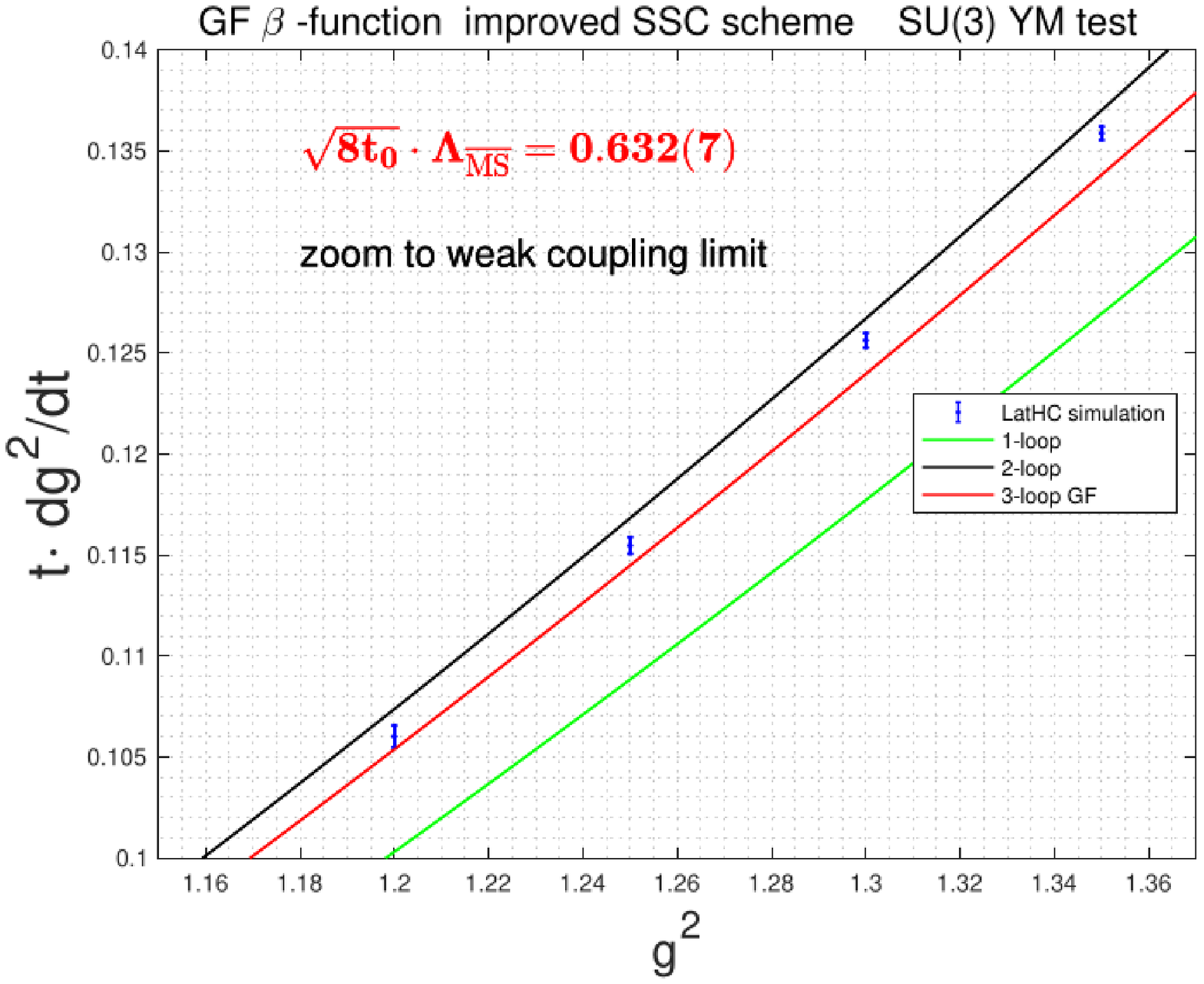}}
        \scalebox{0.45}{\includegraphics{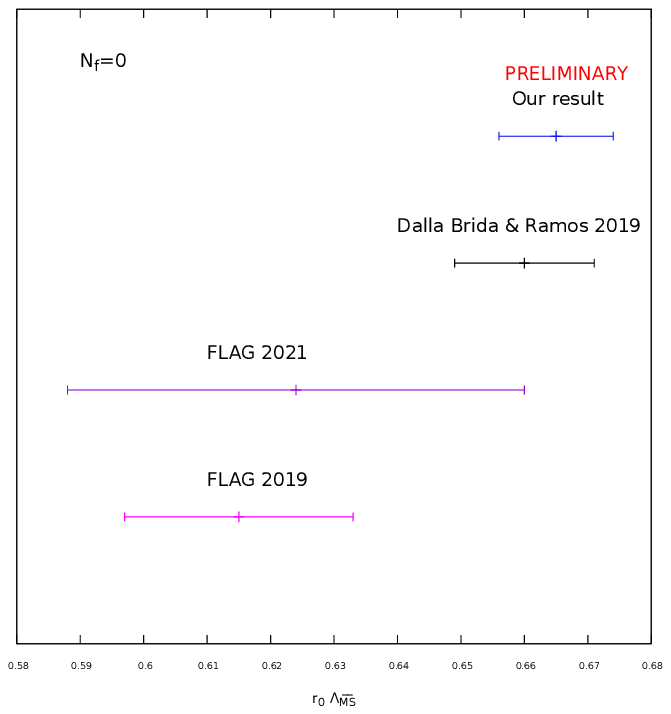}}
	\end{center}		
	\vskip -20pt
    \caption{\label{YM4}  {\small The infinite volume based continuum beta function  $t\cdot dg^2/dt$ is shown in the $g^2 = 1.2-16.4$  range of the renormalized coupling. Precision contact is made matching 3-loop perturbation theory with gradient flow based renormalization at weak coupling. The right panel compares our preliminary fit results with recently reported FLAG reviews.}}
\end{figure}

The three-step procedure can be repeated for any selection of target $g^2$ values in the covered range of the available lattice ensembles. Fig.~\ref{YM4} shows the simulation results for $t\cdot dg^2/dt$ as a function of $g^2$ in the extended $g^2 = 1.2-16.4$ range. Based on the  high precision of $t\cdot dg^2/dt$ expressed as a function of $g^2$ we can calculate  the value of $t_0\cdot  \Lambda_{\overline{\rm MS}}$ in the Yang-Mills theory.  First, $t_0\cdot  \Lambda_{GF}$ is calculated from Eq. (2) with ${\bar g}^2$ set to $g^2(t_0) = 15.79$,
\begin{equation}
t_0\cdot \Lambda_{GF} = (b_0{\bar g}^2)^{-b_1/2b_0^2}\cdot {\rm exp}(-1/2b_0{\bar g}^2)\cdot {\rm  exp}\Bigl (-\int_0^{\bar g} dx\bigl [1/\beta(x) + 1/b_0x^3 - b_1/b_0^2x\bigr] \Bigr ).
\end{equation}
The integral in Eq.(2) was broken up into two parts.
In the $g^2=0-1.2$ range the three-loop value of the $\beta$-function was used and the $g^2=1.2-g^2(t_0)$ range was evaluated with numerical integration, based on  spline fit to the data. The result of $t_0\cdot\Lambda_{GF} = 1.184(13)$ can be converted from a well-know one-loop calculation~\cite{Harlander:2016vzb,Artz:2019bpr} to $t_0\cdot\Lambda_{\overline{\rm MS}}=0.632(7)$, accurate on the percent level.\

Our preliminary result of this pilot study has comparable accuracy to the most recent work which used the finite-volume step  $\beta$-function by authors from the Alpha collaboration~\cite{DallaBrida:2019wur}  with $t_0\cdot\Lambda_{\overline{\rm MS}}=0.6227(98)$ reported and shown in Fig. \ref{YM4}. Results from the infinite volume based $\beta$-function agree within one standard deviation with the result from the Schr\"odinger functional based step $\beta$-function in~\cite{DallaBrida:2019wur}. Our test is very promising although we will subject the preliminary analysis to further scrutiny from continued runs of our lattice ensembles.  

 We converted our result to the $r_0$ scale using $\sqrt{8t_0}/r_0 = 0.948(7)$ from~\cite{Luscher:2010iy}, obtaining $r_0\cdot\Lambda_{\overline{\rm MS}}=0.665(9)$. Connection with the $r_0$ scale was done differently in~\cite{DallaBrida:2019wur} with the result $r_0\cdot\Lambda_{\overline{\rm MS}}=0.660(11)$. Significantly lower world average $r_0\cdot\Lambda_{\overline{\rm MS}}=0.615(18)$ was reported in the FLAG2019 Review which shifted to $r_0\cdot\Lambda_{\overline{\rm MS}}=0.624(36)$ in the  FLAG2021 Review~\cite{Aoki:2021kgd}. The shift of the central value and its increased error came from  including the new result from~\cite{DallaBrida:2019wur} in the  world average. The more than 3-sigma deviation of $r_0\cdot\Lambda_{\overline{\rm MS}}=0.660(11)$ in~\cite{DallaBrida:2019wur} from the central value of the FLAG average without it remains one of the remaining unresolved issues in the Yang-Mills sector~\cite{DallaBrida:2020pag}.
\vskip 0.05in

\section{Precision tests of QCD with ten massless fermions}
\vskip -0.1in
The promising tests of the infinite volume based $\beta$-function of the Yang-Mills model in Section 3 further motivate new efforts  for the determination of $r_0\cdot\Lambda_{\overline{\rm MS}}$ in QCD with three massless flavors following a similar lattice analysis of the infinite volume based $\beta$-function. Given the sizable set of our ten-flavor lattice ensembles at a series of the bare gauge coupling $6/g_0^2$ over an extended range of lattice volumes we will report the first test here with dynamical fermions for QCD with ten massless flavors. The ultimate goal is of course the determination of $r_0\cdot\Lambda_{\overline{\rm MS}}$ with three massless fermions. The original motivation for selecting the ten-flavor model as a pilot study was to show that results from the lattice analysis of the  infinite-volume based ten-flavor $\beta$-function are consistent with the absence of  IRFP from our finite-volume based ten-flavor step $\beta$-function within the controlled lattice reach of renormalized couplings. The analysis of the latter was reported in~\cite{Fodor:2022qfe} with journal publication in preparation.

The tests presented in this section serve as a first pilot study toward the goal of developing our alternate approach to the determination  of the strong coupling $\alpha_s$ at the Z-boson pole in QCD. The growing number of our lattice ensembles  with three massless fermions at a series of the bare gauge coupling $6/g_0^2$ over an extended range of lattice volumes will superseed the limited accuracy presented here for ten flavors but the lattice approach and its analysis will be similar in QCD with three massless flavors.

\begin{figure}[h!]
	\begin{center}
		\scalebox{0.3}{\includegraphics{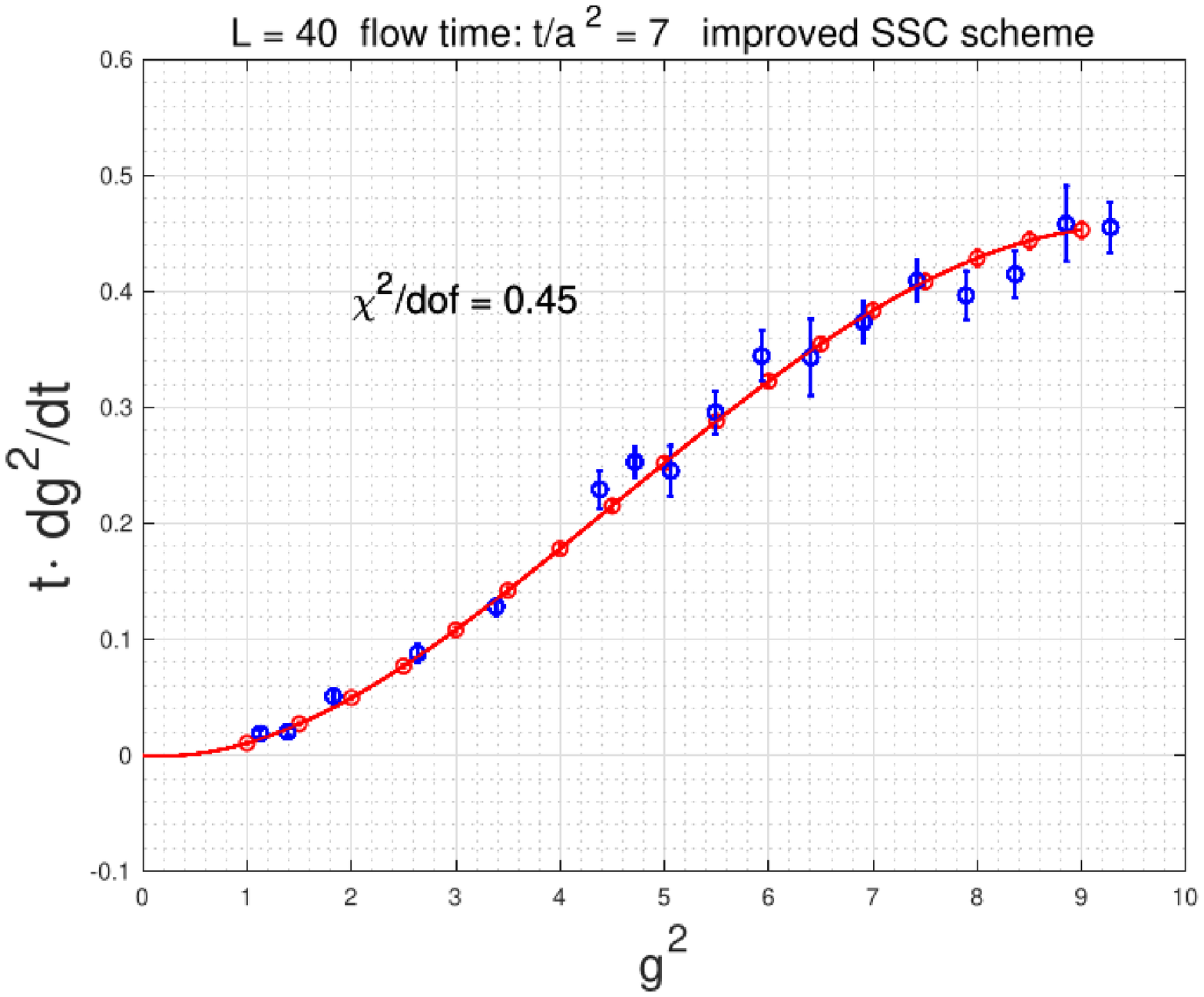}}
		\scalebox{0.3}{\includegraphics{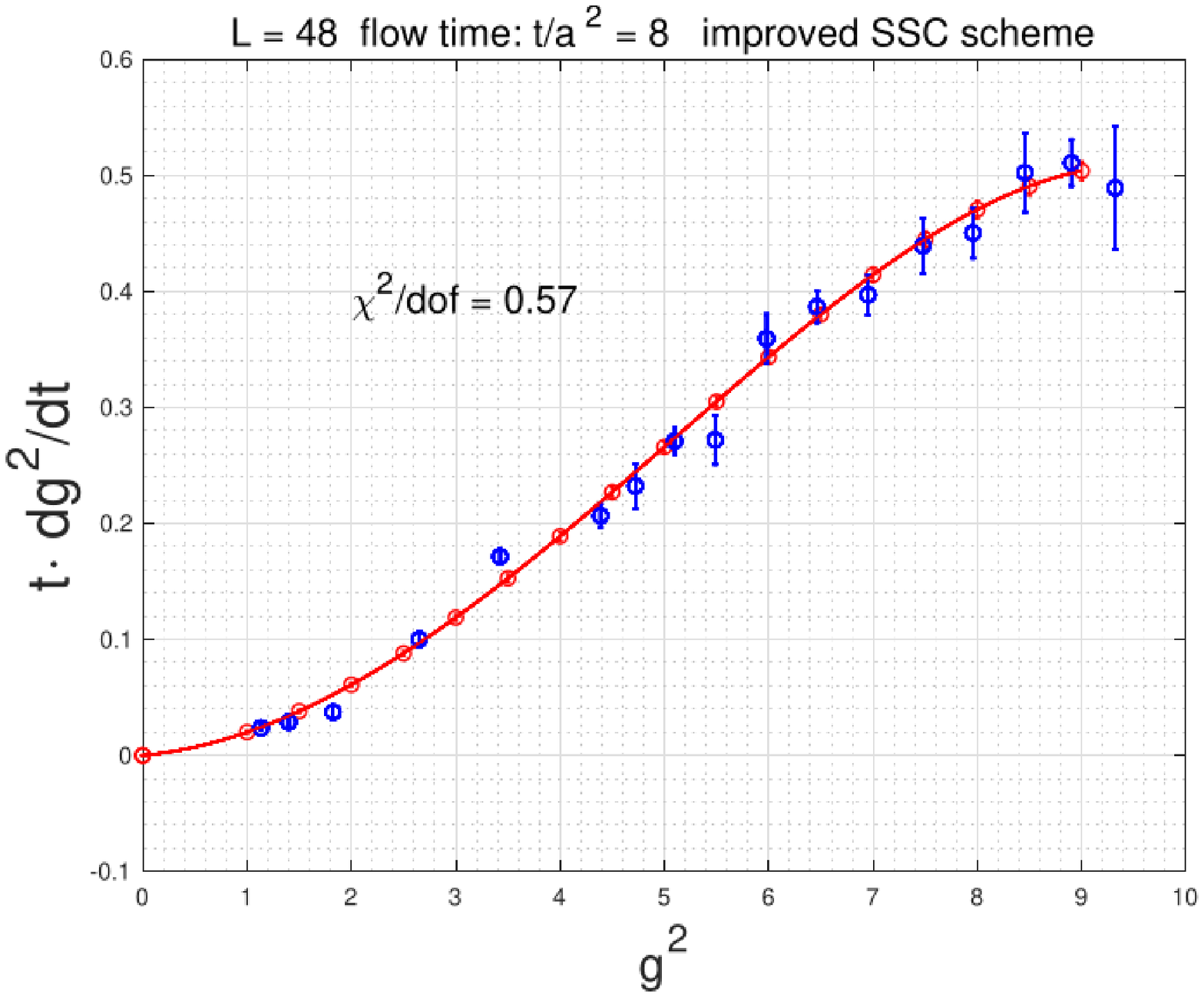}}
	\end{center}		
	\vskip -25pt
	\caption{\label{ContBeta1}  {\small Fourth order polynomial fits of $t\cdot g^2/dt$ are shown in the improved $SSC$ scheme  for $L=40$ and $L=48$
			with good statistical quality. On the left panel the polynomial fit is shown to the blue  data points $t\cdot g^2/dt$ for $L=40$. The interpolkated values are determined from the fit at the targeted  $g^2$  locations for preset flow time $t/a^2 = 7$.  The interpolated values of $t\cdot g^2/dt$  are marked with red symbols for  targeted $g^2$ values.  On the right panel similar fit is shown with interpolated red points for $L=48$ and $t/a^2 = 8$.     }}
	
\end{figure}

One of the most important results in this report is to show  how to make  contact at weak coupling with gradient flow based three-loop perturbation theory in infinite volume~\cite{Harlander:2016vzb,Artz:2019bpr} using the ten-flavor lattice implementation of the infinite volume based $\beta$-function. This contact allows the determination of  $\Lambda_{\overline{\rm MS}}$  in units of a chosen scale $\mu$ implicitely defined by a choice of the renormalized gauge coupling $g^2(\mu=1/\sqrt{8t})$ in the strongly coupled regime. 
There are three steps in the algorithmic implementation of the lattice analysis with somewhat different ordering of the steps compared with what was presented for the Yang-Mills analysis:

\vskip 0.05in
\noindent{\bf Step 1:} For data sets of the analysis, the four  largest volumes $L^4 = 32^4, 36^4, 40^4, 48^4 $  of our ten-flavor lattice ensembles were selected from ~\cite{Fodor:2022qfe} with periodic gauge and antiperiodic fermion boundary conditions at 21  bare gauge couplings $6/g_0^2=2.6,2.7,2.8,2.9,3.0,3.1,3.2,3.3,3.4,3.5,3.6,$ $3.7,3.8,3.9,4.0,4.1,4.5,5.0,6.0,7.0,8.0$.  
For extrapolation to the $a^2/t\rightarrow 0$ continuum limit, we select first twelve targeted flow time values in the range $t/a^2=2.5-8$ in increments of $0.5$. The selected range and the spacings are somewhat arbitrary to balance cutoff effects and finite volume effects in the analysis.  
We also select targeted values of the renormalized coupling in the range $g^2 = 1-10.5$ with increments of $0.5$. The range in $g^2$ is set by the reach of the selected data sets for the analysis  but the spacings are arbitrary. 
For each $L$ and for each selected $t/a^2$ we interpolate $t\cdot dg^2/dt$  to the selected gauge coupling $g^2$  which is held fixed in the three-step process. At each target gauge coupling $g^2$ the interpolated value of  $t\cdot dg^2/dt$  is obtained for each preset value of $t/a^2$ using fourth order polynomial fit as a function of  $g^2$.  Some fits are shown in Fig.~\ref{ContBeta1}.

\begin{figure}[h!]
	\begin{center}
		\scalebox{0.3}{\includegraphics{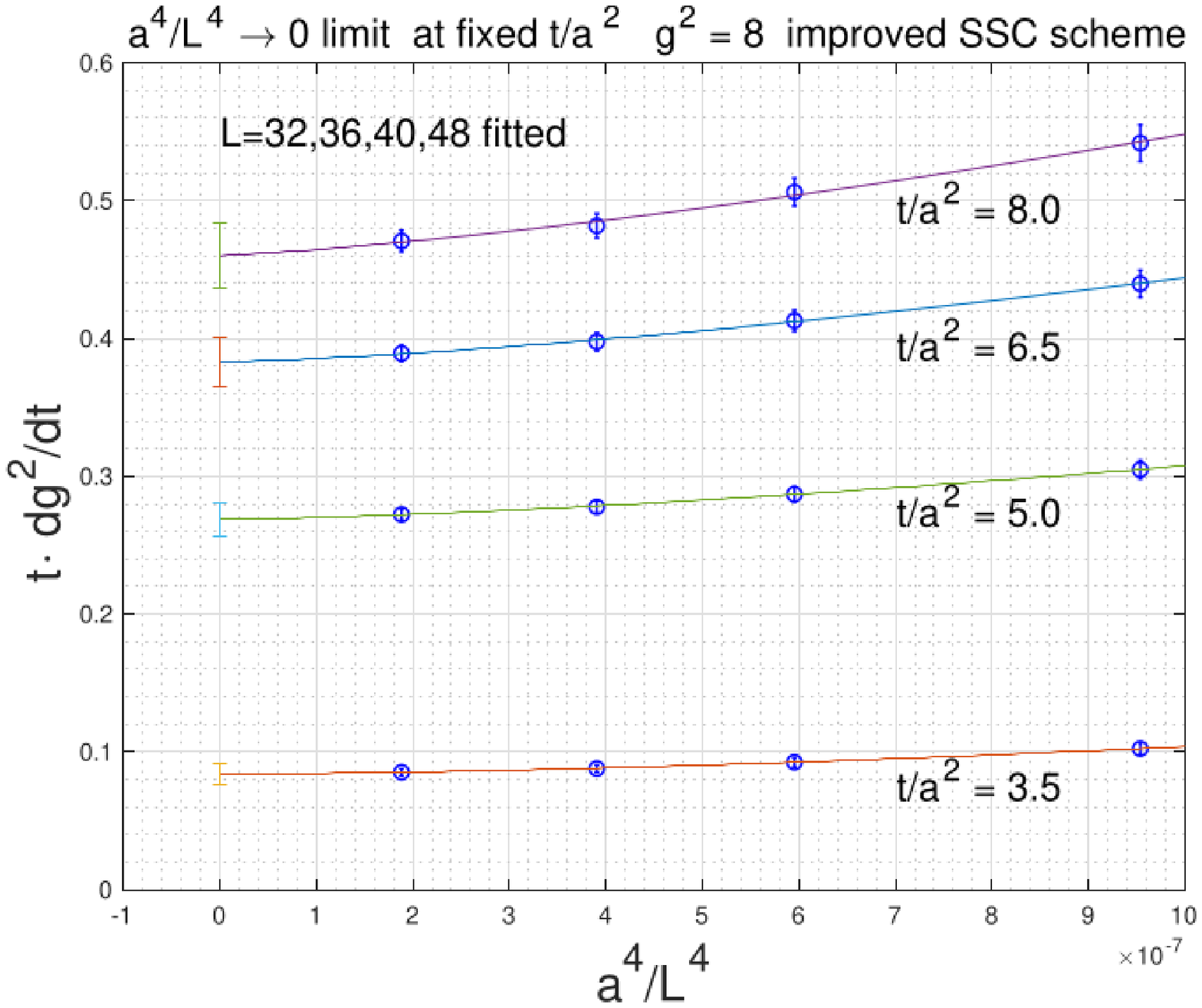}}
		\scalebox{0.3}{\includegraphics{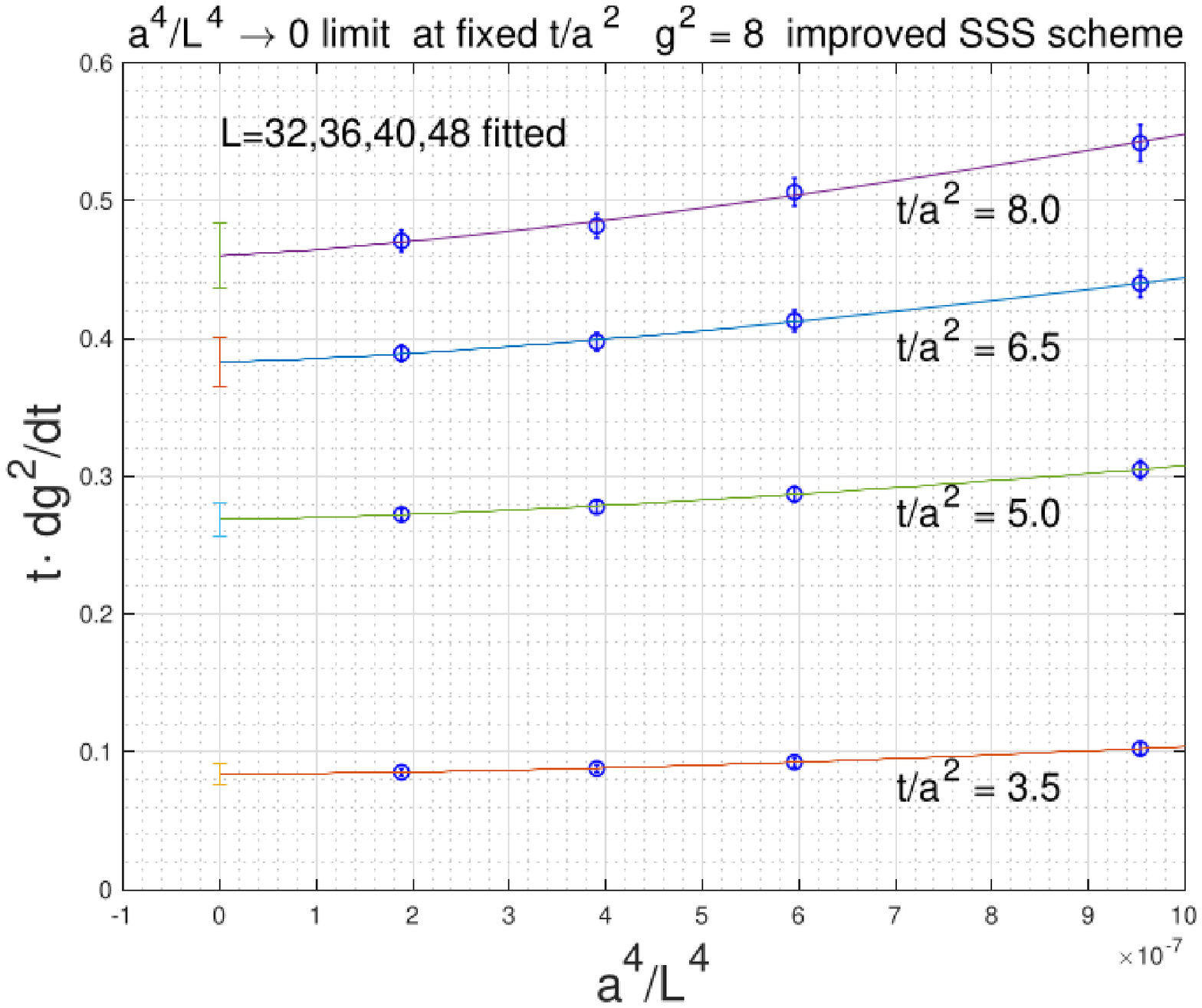}}
	\end{center}		
	\vskip -20pt
	\caption{\label{ContBeta2}  {\small  The interpolated $L$-dependent beta-functions of $\beta=t\cdot dg^2/dt$ are fitted  in the $a^2/L^4 \rightarrow 0$  infinite volume limit 
			at fixed lattice spacing set by four representative values of  $t/a^2$ from the $2.5-8$ range at $g^2=8$.  The fits with very good statistics are polynomials in the $a^2/L^2$ 
			variable with a leading $a^4/L^4$ term and the higher order $a^6/L^6$ term with $L/a=32,36,40,48$ in the improved SSC scheme on the left and the improved SSS scheme on the right. }}
\end{figure}

\begin{figure}[h!]
	\begin{center}
		\scalebox{0.3}{\includegraphics{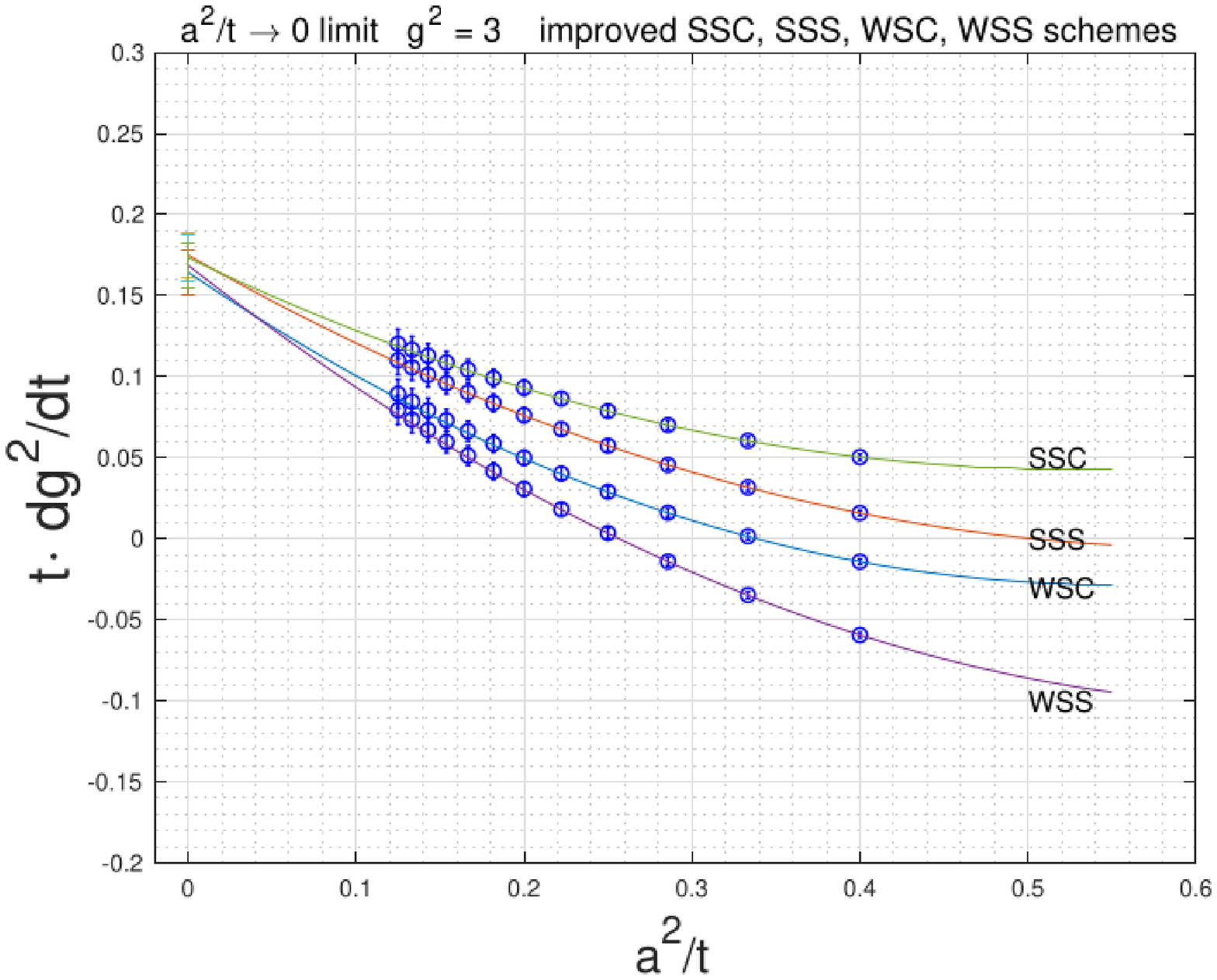}}
		\scalebox{0.3}{\includegraphics{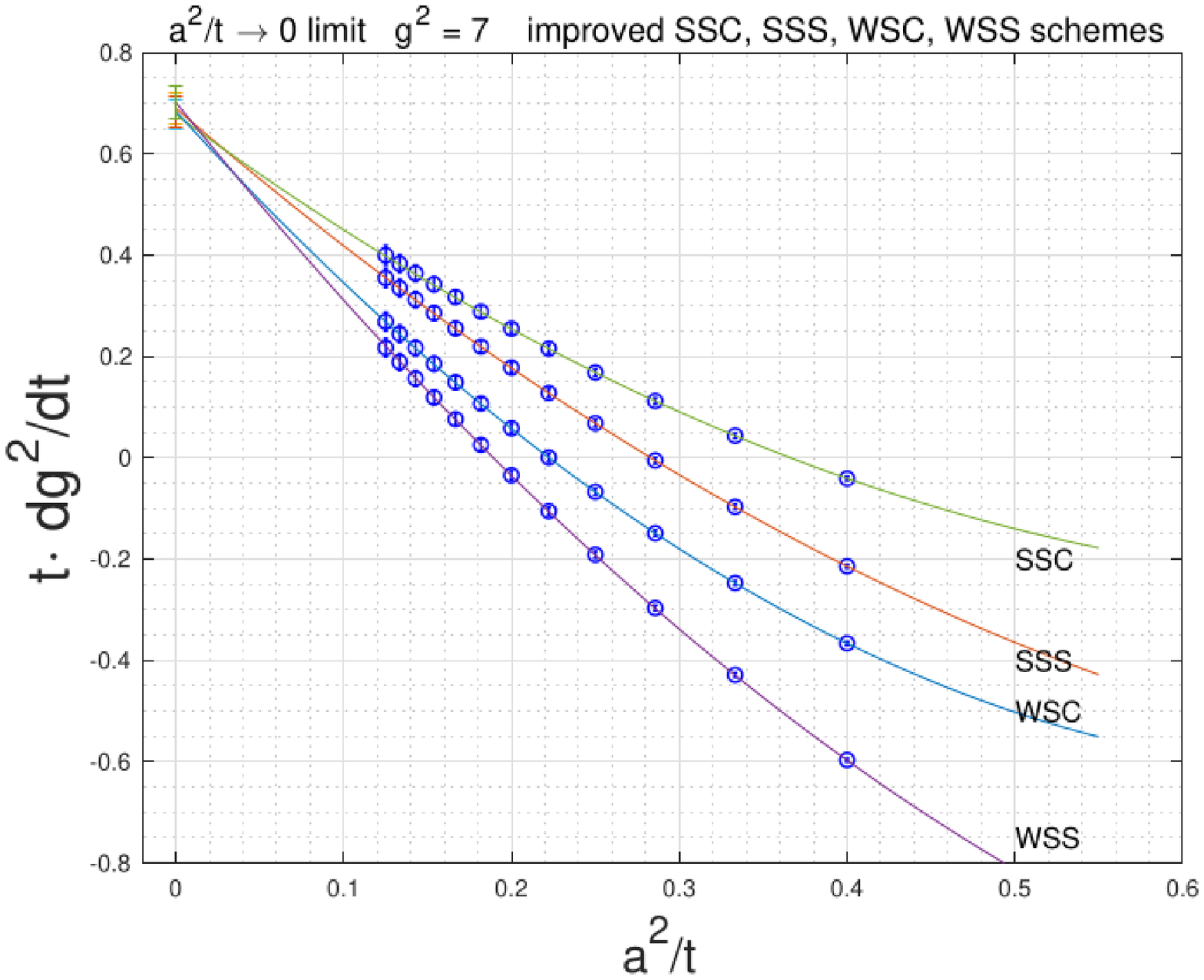}}
	\end{center}		
	\vskip -20pt
	\caption{\label{ContBeta3}  {\small Continuum fits of $\beta=t\cdot dg^2/dt$ are shown in all four schemes at two renormalized couplings with choices of $g^2=3$
			and $g^2=7$. The fits are quadratic in the $a^2/t$ variable.  }}
	
\end{figure}

\vskip 0.05in
\noindent{\bf Step 2:}  The interpolated $L$-dependent $\beta$-functions $t\cdot dg^2/dt$  are extrapolated to the infinite volume limit for each $t/a^2$ at the target $g^2$ from Step 1, with samples of these fits shown in Fig~\ref{ContBeta2} in the variable $a^4/L^4$.
\newpage
\vskip -0.05in
\noindent{\bf Step 3:}  After Step 2 we have the infinite volume based $\beta$-functions $t\cdot dg^2/dt$ at 12 values of $t/a^2$ for each targeted value of the gradient flow based renormalized coupling $g^2$ held fixed in the algorithm. In Step 3, we fit the cutoff dependence of   $t\cdot dg^2/dt$  to determine its $a^2/t\rightarrow 0$ continuum limit at fixed $g^2$. This is obtained by quadratic fits in the $a^2/t$ variable with fit samples shown in Fig.~\ref{ContBeta3}.

\begin{figure}[h!]
	\begin{center}
		\scalebox{0.25}{\includegraphics{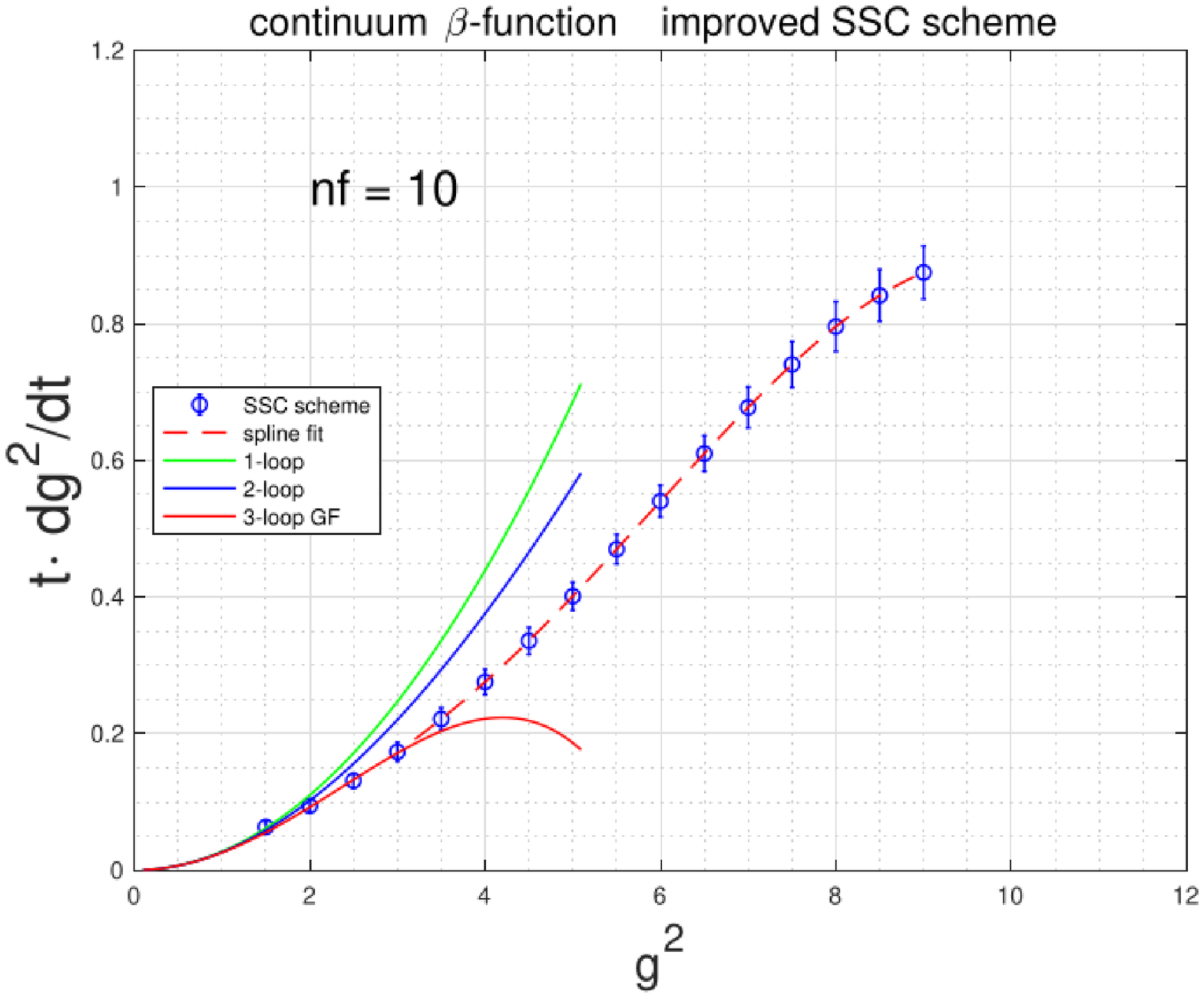}}
		\scalebox{0.25}{\includegraphics{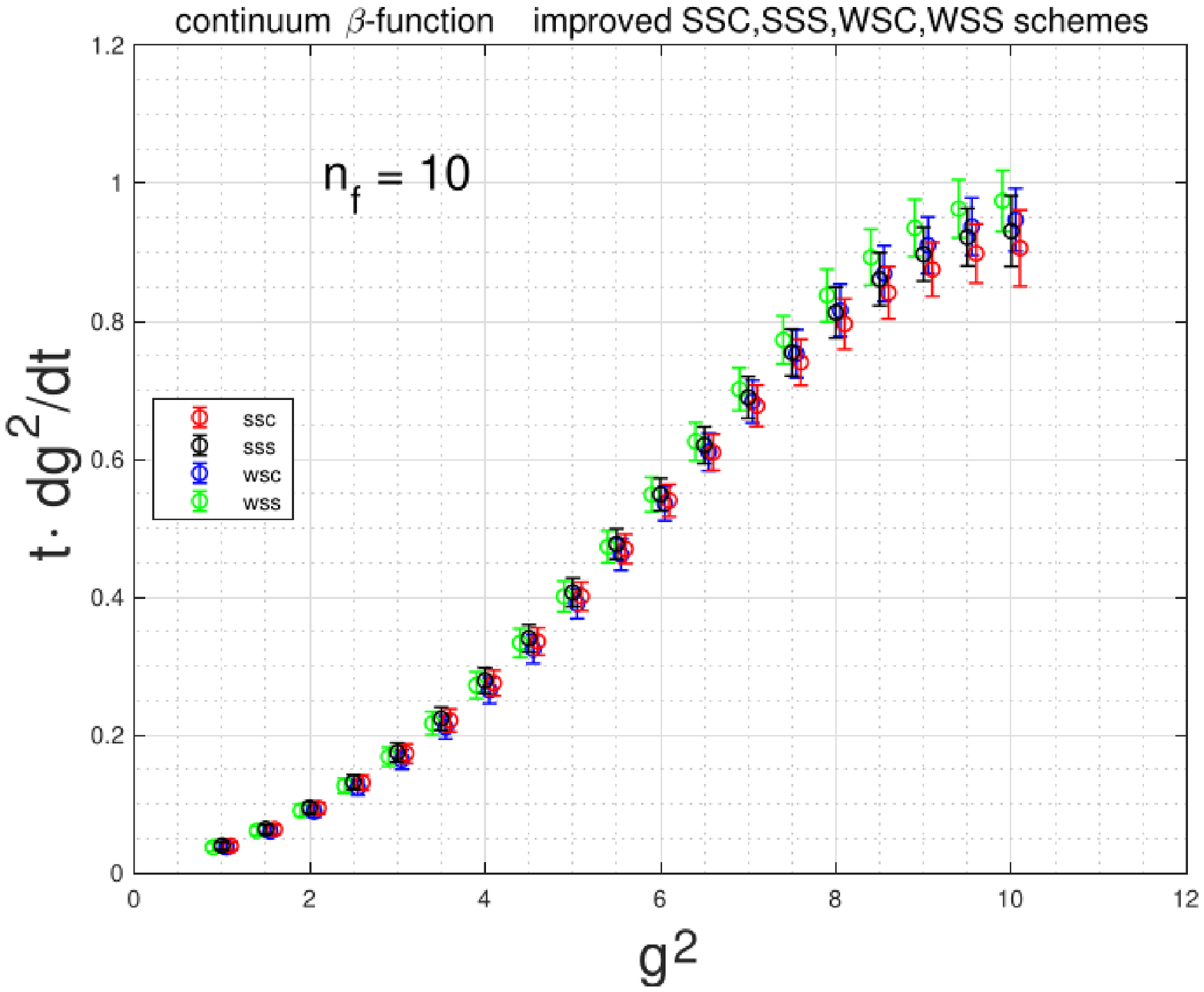}}
		\scalebox{0.25}{\includegraphics{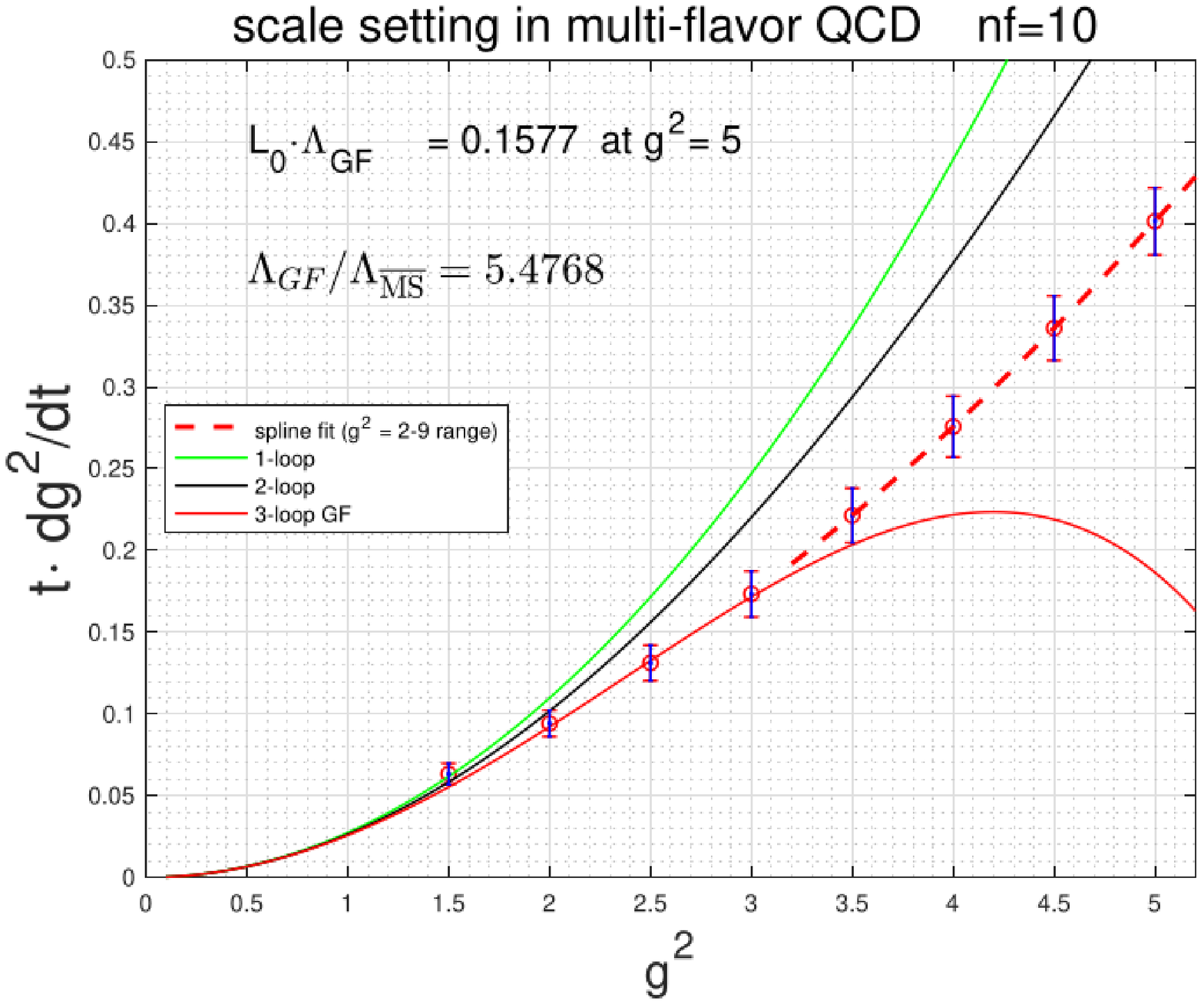}}
	\end{center}		
	\vskip -20pt
	\caption{\label{ContBeta4}  {\small The dashed line in the left panel shows spline based fitting in the  $g^2=3-9$  range of the SSC scheme. The fit would be identical in the SSS scheme, or in the WSC and WSS  schemes,  added  in the middle panel. Below $g^2=3$ three-loop perturbation theory can be used for the $\beta$-function from~\cite{Harlander:2016vzb,Artz:2019bpr}, like shown in the right panel. }}
\end{figure}

\newpage
With the three-step analysis of the ten-flavor model we reached the set goal of the infinite volume based $\beta$-function  given by $t\cdot dg^2/dt$ in the $g^2(t)  = 1.5-10.5$ range, as shown in Fig.~\ref{ContBeta4}. The definitions of the improved SSS, SSC, WSC, and WSS  schemes of the 3-step procedure are given in~\cite{Fodor:2022qfe}. 

Similar to~\cite{DallaBrida:2018rfy}  in three-flavor QCD with massless fermions, we can connect now the $\Lambda_{GF}$ scale of the gradient flow scheme to other scales, like the scale $\mu=1/\sqrt{8t}$ set by the choice $g^2(t)$ in the ten-flavor theory with massless fermions at some flow time $t$. 
As an example, we will express the scale $L_0 = \sqrt{8t}$ set implicitly at ${\bar g}^2 (t)= 5$ in $\Lambda_{GF}$ units of the ten-flavor theory,
\begin{equation}\label{lambda_10}
	L_0\cdot \Lambda_{GF} = (b_0{\bar g}^2)^{-b_1/2b_0^2}\cdot {\rm exp}(-1/2b_0{\bar g}^2)\cdot {\rm  exp}\Bigl (-\int_0^{\bar g} dx\bigl [1/\beta(x) + 1/b_0x^3 - b_1/b_0^2x\bigr] \Bigr ).
\end{equation}
The integral in Eq.~(\ref{lambda_10}) can be broken up into two parts. In the $x=0-3$ range the three-loop value of the $\beta$-function was used and the $x=3-5$ range was evaluated with numerical  integration, based on spline fits to the data. The result of $L_0\cdot\Lambda_{GF} = 0.1577$ can be converted to $\Lambda_{\overline{\rm MS}}$  units of the ten-flavor theory,
$L_0\cdot  \Lambda_{\overline{\rm MS}} = 0.02879$, using the conversion 
factor $\Lambda_{GF}/\Lambda_{\overline{\rm MS}}=5.4768$ from a well-know one-loop calculation~\cite{Harlander:2016vzb,Artz:2019bpr}.
The precision of the calculation is on the few percent level with combined systematic and statistical uncertainties, not very far from the challenging goal of performing similar analysis in three-flavor QCD with massless fermions on the 1-2 percent level using $t_0$ or $r_0$ for scale setting~\cite{Luscher:2010iy}.

\noindent{\bf Conclusions:}
We have shown evidence for the feasibility of our infinite volume based step $\beta$-function analysis in the Yang-Mills sector of quenched QCD and in massless multi-flavor QCD  with ten flavors. We determined $r_0\cdot\Lambda_{\overline{\rm MS}}=0.665(9)$ in the Yang-Mills sector and  successfully expressed $\Lambda_{\overline{\rm MS}}$ of the ten-flavor theory in terms of a scale set implicitly at flow time $t$ by $g^2(t)=5$  in the ten-flavor test. 
As illustrated in Fig.~\ref{ContBeta4},  the challenge is to apply similar analysis to the strong coupling $\alpha_s$ in QCD with three massless flavors.
To reach the aimed 1-2 percent accuracy in three-flavor QCD requires carefully selected broad range of lattice ansembles in the barge gauge coupling $6/g_0^2$, with large lattice sizes, like  $L=64$, and long runs of the ensembles in the $10K-20K$ trajectory range, as in our ongoing investigations.
\vskip -0.1in
\begin{acknowledgments}
	\vskip -0.1in
	We acknowledge support by the DOE under grant DE-SC0009919, by the NSF under grant 1620845,  and by the Deutsche Forschungsgemeinschaft grant SFB-TR 55. Computations for this work were carried out in part on facilities of the USQCD Collaboration, which are funded by the Office of Science of the U.S. Department of Energy. Computational resources were also provided by the DOE INCITE program  on the SUMMIT gpu platform at ORNL, by the University of Wuppertal, and by the Juelich Supercomputing Center. 
\end{acknowledgments}

\end{document}